\documentclass[twocolumn,showpacs,preprintnumbers,amsmath,amssymb]{revtex4}

\usepackage{graphicx}
\usepackage{dcolumn}
\usepackage{bm}

\begin{document}

\preprint{APS/123-QED}

\title{Investigations of the $\pi N$ total cross sections at high energies 
using\\ new FESR:   log $\nu$ or (log $\nu )^2$  }

\author{Keiji  Igi}
 \email{igi@info.kanagawa-u.ac.jp}
\affiliation{%
Department of Information Science, Kanagawa University, Hiratsuka\\
Kanagawa 259-1293, Japan
}%

\author{Muneyuki Ishida}
 \email{ishida@th.phys.titech.ac.jp}
\affiliation{
Department of Physics, Tokyo Institute of Technology\\
Tokyo 152-8551, Japan
}%

\date{\today}

\begin{abstract}
We propose to use rich informations on $\pi p$ total cross sections below 
$N(\sim 10$ GeV) in addition to high-energy data
in order to discriminate whether these cross sections increase
like log $\nu$ or (log $\nu )^2$ at high energies,
since it is difficult to discriminate between asymptotic log $\nu$ and (log $\nu )^2$
fits from high-energy data alone.
 A finite-energy sum rule (FESR)
which is derived in the spirit of the $P^\prime$ sum rule as well as the $n=1$ moment
FESR have been required to constrain the high-energy parameters. We then searched
for the best fit of $\sigma_{\rm tot}^{(+)}$ above 70 GeV 
in terms of high-energy parameters
constrained by these two FESR. 
We can show from this analysis that the (log $\nu )^2$ behaviours is preferred
to the log $\nu$ behaviours.  
\end{abstract}

\pacs{13.60.Hb   12.40.Nn}
\maketitle


The high energy behavior of $\pi N$ total cross sections has been
one of the longstanding
problems in particle physics. 
The sum of $\pi^+ p$ and $\pi^- p$ total 
cross sections has a tendency to increase above 70 GeV 
experimentally\cite{rf1}. 
It is well-known as the Froissart unitarity bound\cite{rf2}
that the increase of total cross sections is at most log$^2\nu$.
It has not been possible\cite{Cudell}, however, 
to discriminate between asymptotic log $\nu$ and log$^2\nu$ fits if one uses
high-energy data alone above $\sim$70GeV.

The purpose of this paper is to propose to use rich informations of  $\pi p$
total cross sections at low and intermediate energy regions
{\it through new finite-energy sum rules (FESR) as constraints}
in addition to high-energy data, in order to 
discriminate the high energy behaviours of $\pi p$ total cross sections 
above 70 GeV. 
   
Such a kind of attempt has been initiated in Ref.~\cite{rf3}.
The $s$-wave $\pi N$ scattering length $a^{(+)}$ of the crossing-even amplitude had been
expressed as 
\begin{eqnarray}
\left(  1+\frac{\mu}{M}  \right) a^{(+)} &=& -\frac{g_r^2}{4\pi}\left( \frac{\mu}{2M} \right)^2
\frac{1}{M} \frac{1}{1-(\frac{\mu}{2M})^2} \nonumber\\
& & +\frac{1}{2\pi^2}\int_0^\infty dk 
[\sigma_{\rm tot}^{(+)}(k)-\sigma_{\rm tot}^{(+)}(\infty )]\ \ \ \ 
\label{eq1}
\end{eqnarray}
with pion mass $\mu$ under the assumption that there are no singularities 
with the vacuum quantum numbers in the $J$ plane except for the Pomeron $(P)$.
The evidence that this sum rule had not been satisfied led us to the prediction of the 
$P^\prime$ trajectory with $\alpha_{P^\prime}(0)\approx 0.5$, and 
soon the $f$ meson ($f_2(1275)$)
has been uncovered on this $P^\prime$ trajectory.\\

\hspace*{-0.4cm}(\underline{\it FESR(1)}): Taking into account 
the present situation of increasing total cross section data,
we derive FESR in the spirit of the $P^\prime$ sum rule\cite{rf3}. 
We consider the 
crossing-even (spin-averaged) forward scattering amplitude for $\pi p$ scattering\cite{rf4} 
\begin{eqnarray}
f^{(+)}(\nu ) &=& \frac{1}{4\pi} [ A^{(+)}(\nu )+\nu B^{(+)}(\nu ) ] .
\label{eq2}
\end{eqnarray}
We assume
\begin{eqnarray}
{\rm Im}\  f^{(+)}(\nu )  & \simeq & {\rm Im}\ R(\nu )
+{\rm Im}\ f_{P^\prime}(\nu ) \nonumber\\
 = \frac{\nu}{\mu^2}  ( c_0 &+& c_1 {\rm log}\ \frac{\nu}{\mu} 
+ c_2 {\rm log}^2\ \frac{\nu}{\mu}    ) 
+\frac{\beta_{P^\prime}}{\mu} 
\left( \frac{\nu}{\mu} \right)^{\alpha_{P^\prime}(0)}
\label{eq3}
\end{eqnarray}
at high energies $(\nu \geq N)$. 
Since this amplitude is crossing-even, we have
\begin{eqnarray}
R(\nu ) &=& \frac{i\nu}{2\mu^2} \{ 2c_0+c_2\pi^2
+c_1\left( {\rm log}\frac{e^{-i\pi}\nu}{\mu}
+{\rm log}\frac{\nu}{\mu} \right) \nonumber\\
&& +c_2\left( {\rm log}^2\frac{e^{-i\pi}\nu}{\mu}+{\rm log}^2
\frac{\nu}{\mu} \right)       
\}, \ \ \ \ \ \ \    \\
f_{P^\prime}(\nu ) &=& -\frac{\beta_{P^\prime}}{\mu}\left(   
\frac{  ( e^{-i\pi}\nu /\mu )^{\alpha_{P^\prime}(0)}
+(\nu /\mu )^{\alpha_{P^\prime}(0)} }{
{\rm sin}\ \pi \alpha_{P^\prime}(0)}    \right) , 
\label{eq55}
\end{eqnarray}
and subsequently we obtain 
\begin{eqnarray}
{\rm Re}\ R(\nu) &=& \frac{\pi\nu}{2\mu^2} \left( c_1+2c_2 {\rm log}\ 
\frac{\nu}{\mu}  \right) \ ,
\label{eq4} \\ 
{\rm Re}\ f_{P^\prime}(\nu) &=& 
-\frac{\beta_{P^\prime}}{\mu} \left(  \frac{\nu}{\mu}  \right)^{\alpha_{P^\prime}(0)}
{\rm cot} \frac{\pi \alpha_{P^\prime}(0)}{2} \nonumber\\
 &=&
-\frac{\beta_{P^\prime}}{\mu} \left(  \frac{\nu}{\mu}  \right)^{0.5} \ ,
\label{eq5}
\end{eqnarray}
substituting $\alpha_{P^\prime}(0)=\frac{1}{2}$ in Eq.~(\ref{eq55}).
Let us define
\begin{eqnarray}
\tilde f^{(+)} (\nu) &=& f^{(+)}(\nu ) -R(\nu )-f_{P^\prime}(\nu ) \nonumber\\
&\sim& \nu^{\alpha (0)}\ \ \  (\alpha (0)<0),
\label{eq6}
\end{eqnarray}
and write dispersion relation for $\frac{\tilde f^{(+)}(\nu )}{\nu -\mu }$.
Since this amplitude is superconvergent, we obtain 
\begin{eqnarray}
{\rm Re}\ \tilde f^{(+)}(\mu ) &=& 
\frac{P}{\pi} \int_{-\infty}^{\infty} d\nu^\prime 
\frac{{\rm Im}\ \tilde f^{(+)}(\nu^\prime )}{\nu^\prime -\mu } \nonumber\\
&=& \frac{2P}{\pi} \int_{0}^{\infty} 
\frac{\nu^\prime{\rm Im}\ \tilde f^{(+)}(\nu^\prime )}{k^{\prime 2} }
d\nu^\prime  \ .
\label{eq7}
\end{eqnarray}
Using Eqs.~(\ref{eq6}) and (\ref{eq7}), we have 
\begin{widetext}
\begin{eqnarray}
{\rm Re}\ f^{(+)}(\mu ) &=& {\rm Re}\ R(\mu) + {\rm Re}\ f_{P^\prime}(\mu) 
-\frac{g_r^2}{4\pi}\left(\frac{\mu}{2M}\right)^2\frac{1}{M}\frac{1}{1-(\frac{\mu}{2M})^2} \nonumber\\
&&+\frac{1}{2\pi^2}\int_0^{\overline{N}} \sigma_{\rm tot}^{(+)}(k) dk
-\frac{2P}{\pi}\int_0^N \frac{\nu}{k^2}
\left\{ 
{\rm Im}\ R(\nu )+\frac{\beta_{P^\prime}}{\mu}
\left(\frac{\nu}{\mu}\right)^{0.5}  \right\} d\nu \ ,\ \ \ \ \ \ \ \ \ \ \ 
\label{eq8} 
\end{eqnarray}
\end{widetext}
where  $\overline{N}\equiv \sqrt{N^2-\mu^2}\simeq N$. 
Let us call Eq.~(\ref{eq8}) as the FESR(1) which we use as the first constraint.
It is important to notice that Eq.~(\ref{eq8}) reduces to the $P^\prime$ sum rule
in ref.\cite{rf3} if $c_1,\ c_2\rightarrow 0$.

The FESR (\cite{rf5}, \cite{rf6}, \cite{rf7})
\begin{eqnarray}
\int_0^N d\nu \ \nu^n {\rm Im}\ f(\nu ) &=& \sum_i \beta_i \frac{N^{\alpha_i(0)+n+1}}{\alpha_i(0)+n+1}
\label{eq88}
\end{eqnarray}
holds for even positive integer $n$ when $f(\nu )$ is crossing odd,
and holds for odd positive integer $n$ when $f(\nu )$ is crossing even.
We can also derive negative-integer moment FESR.
The only significant FESR is a one for $f^{(+)}(\nu )/\nu $ corresponding to $n=-1$.
FESR(1) belongs to this case.

It is important to emphasize that the FESR should not depend so much 
on the value of $N$.\\

\hspace*{-0.4cm}(\underline{\it FESR(2)}): 
The second FESR corresponding to $n=1$ is:
\begin{widetext}
\begin{eqnarray} 
\pi\mu \left( \frac{g_r^2}{4\pi} \right) \left( \frac{\mu}{2M} \right)^3
 &+&  \frac{1}{4\pi}\int_0^{\overline{N}}dk\ k^2\sigma_{\rm tot}^{(+)}(k)
 = \int_0^N \nu {\rm Im}\ R(\nu ) d\nu + \int_0^N \nu {\rm Im}\ f_{P^\prime}(\nu )d\nu\ \ . 
\label{eq9}
\end{eqnarray}
\end{widetext}
We call Eq.~(\ref{eq9}) as the FESR(2). It is to be noticed that the contribution from
higher energy regions is enhanced. \\

\hspace*{-0.4cm}(\underline{\it Data})\ \ The numerical values,
\begin{eqnarray} 
-\frac{g_r^2}{4\pi}\left(\frac{\mu}{2M}\right)^2
\frac{1}{M}\frac{1}{1-(\frac{\mu}{2M})^2}
 &=& -0.0854{\rm GeV}^{-1}, \nonumber\\
\pi \mu \frac{g_r^2}{4\pi} \left( \frac{\mu}{2M} \right)^3
 &=& 0.0026{\rm GeV}\ ,
\end{eqnarray}
have been evaluated using $\frac{g_r^2}{4\pi}=14.4$.
\begin{eqnarray}
{\rm Re}\  f^{(+)}(\mu ) &=& \left( 1+\frac{\mu}{M} \right) a^{(+)}
= \left( 1+\frac{\mu}{M} \right) 
\frac{1}{3}(a_{\frac{1}{2}}+2a_{\frac{3}{2}}) \nonumber\\
&=& -(0.014\pm 0.026){\rm GeV}^{-1}
\end{eqnarray} 
was obtained from\cite{rf8} $a_{\frac{1}{2}}=(0.171\pm 0.005)\mu^{-1}$
and $a_{\frac{3}{2}}=-(0.088\pm 0.004)\mu^{-1}$.

We have used rich data\cite{rf1} of $\sigma^{\pi^+ p}$ and $\sigma^{\pi^- p}$
to evaluate the relevant integrals of cross sections
appearing in FESR(1) and (2). 
We connect each data point\footnote{
We take the error $\Delta y$ for each data point $y$ as 
$\Delta y=\sqrt{(\Delta y)_{\rm stat.}^2+(\Delta y)_{\rm syst.}^2}$.
When several data points, denoted as
$y_i$ with error $\Delta y_i$ $(i=1,\cdots ,n)$, are listed at the same value 
of $k$, these points are replaced by $\overline{y}$ with $\Delta\overline{y}$,
given by $\overline{y} =(\sum_i y_i/(\Delta y_i)^2)/(\sum_i 1/(\Delta y_i)^2)$
and $\Delta \overline{y} =\sqrt{1/(\sum_i 1/(\Delta y_i)^2)}$.
}
of $\sigma^{\pi^\pm p}(k)$ or $k^2\sigma^{\pi^\pm p}(k)$ 
with the next point by straight line in order, from $k=0$ to $k=\overline{N}$,
and regard the area of this polygonal line graph as the relevant integral
in the region $0\leq k\leq \overline{N}$. 
The integrals of $\sigma^{(+)}_{\rm tot}(k)$ ($k^2\sigma^{(+)}_{\rm tot}(k)$) 
are given by averaging these of  $\sigma^{\pi^+ p}(k)$ and 
$\sigma^{\pi^- p}(k)$
 ($k^2\sigma^{\pi^+ p}(k)$ and $k^2\sigma^{\pi^- p}(k)$).
%
We have obtained 
\begin{eqnarray}
\frac{1}{2\pi^2}\int_0^{\overline{N}} dk\ 
\sigma_{\rm tot}^{(+)}(k)  &=& 38.75\pm0.25\ {\rm GeV}^{-1},\nonumber\\ 
\frac{1}{4\pi}\int_0^{\overline{N}} dk\ k^2 
\sigma_{\rm tot}^{(+)}(k) &=& 1817\pm31\ {\rm GeV}
\end{eqnarray}
for $\overline{N}=10$ GeV. 
The errors of relevant integrals, which are from the error of each data point,
are very small ($\sim 1$ percent), and thus, we regard the central values 
are exact ones in the following analysis.

When $\sigma^{\pi^+ p}$ and $\sigma^{\pi^- p}$ data points are 
listed at the same value of $k$, 
we make $\sigma^{(+)}_{\rm tot}(k)$ data point 
by averaging these values.
Totally 183 points are obtained in the region $0.16\leq k \leq 340$ GeV 
as $\sigma^{(+)}_{\rm tot}(k)$ data.
There are 12 points in $k\geq 70$GeV region,
which will be used in the following analysis.\\

\hspace*{-0.4cm}(\underline{\it Analysis})\ \  The FESR(1) and (2) are our 
starting points. Armed with these two, we expressed high-energy parameters 
$c_0$, $c_1$, $c_2$, 
$\beta_{P^\prime}$ in terms of the Born term and the $\pi N$ scattering length $a^{(+)}$
as well as the total cross sections up to $N$. We then attempt to 
fit the $\sigma_{\rm tot}^{(+)}$ above 70GeV. We set $N=10GeV$ (corresponding to
$\sqrt{s_{p\pi}}=4.43$GeV) 
since there are no resonances above this energy.
The FESR(2) has also contributions from the lower trajectory $P"$ which may
pass through $f_2(1810)$. Since $\alpha_{P"}(0)$ is expected to be around -1, we can
assume $P"$ contribution to be suppressed compared with that from $P^\prime$.

Let us first define the log$^2\nu$ model and the log $\nu$ model.
The log$^2\nu$ model is a model for which the imaginary part of $f^{(+)}(\nu )$
behaves as $a+b\ {\rm log}\ \nu+c({\rm log}\ \nu)^2$ 
as $\nu$ becomes large\cite{KN75}.
The log $\nu$ model is a model for which the imaginary part of $f^{(+)}(\nu )$
behaves as $a^\prime +b^\prime\ {\rm log}\ \nu$ for large $\nu$.
So we generally assume that the Im $f^{(+)}(\nu )$ behaves as Eq.~(\ref{eq3})
at high energies $(\nu\geq N)$.\\
{\bf (1) log $\nu$ model}:\ \ This model has three parameters 
$c_0$, $c_1$ and $\beta_{P^\prime}$ with two constraints FESR (1), (2).
(Note that the number of independent parameters is one.)
We set $N=10$GeV and expressed
both $c_0$, $\beta_{P^\prime}$ as a function of $c_1$ using the FESR(1) and (2).
We obtained 
\begin{eqnarray}
c_0(c_1) &=& 0.0879-4.94c_1,\nonumber\\ 
\beta_{P^\prime}(c_1) &=& 0.1290+7.06c_1\ .
\label{eq12}
\end{eqnarray}  
We then tried to fit 12 data points of $\sigma_{\rm tot}^{(+)}(k)$ 
between 70GeV and 340GeV. 
The result is shown by thick solid line in FIG. 1.
The best fit we obtained is $c_1=0.00185$ 
which gives $c_0=0.0787$ and $\beta_{P^\prime}=0.142$ with the bad 
``reduced $\chi^2$," 
$\chi^2/(N_{\rm data}-N_{\rm param})=29.04/(12-1)\simeq 2.6$.
Therefore it turned out that this model has difficulties to reproduce 
the experimental increase
of $\pi p$ total cross sections above 70GeV (see, thick solid line in FIG. 1 (b)).
In this log $\nu$ fit, the results also depend on the value of $N$,
which is not so good.

\begin{figure*}
\includegraphics{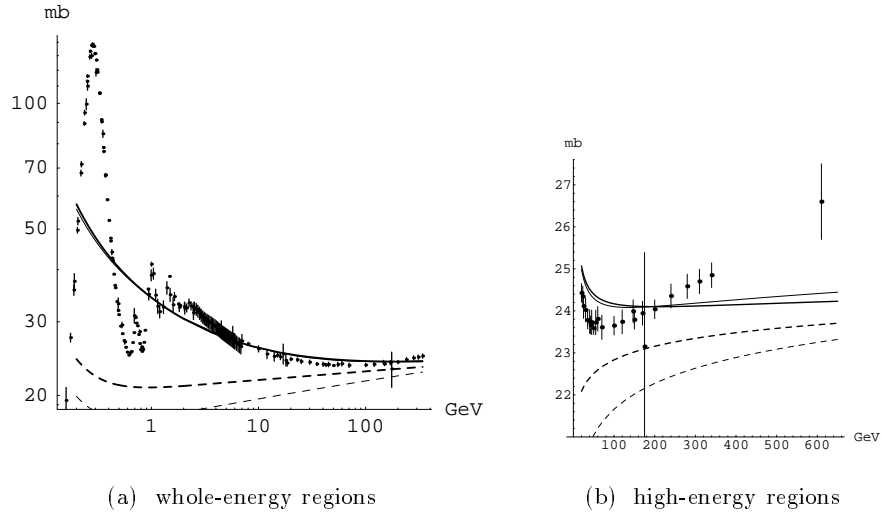}
\caption{\label{fig:wide}Fit to the $\sigma_{\rm tot}^{(+)}$ data above 70GeV 
by the log $\nu$ model.
Thick(Thin) solid line shows the result 
in the case of $\alpha_{P^\prime}=0.5(0.586)$.
Correspondingly, the contribution from Im $R(\nu )$ (with $c_2=0$)
is shown by thick(thin) dashed line.
Recently a datum\cite{rf12} for $\pi^- N$ total cross section 
at very high energy ($k$=610GeV) was reported by the SELEX collaboration.
This point is included in (b). 
The log $\nu$ model with $\alpha_{P^\prime}(0)=0.5(0.586)$
predicts 24.2(24.4)mb for $\sigma_{\rm tot}^{(+)}$ at 610GeV
which is inconsistent with their value on $\pi^- N$, 
$(26.6\pm 0.9)$mb .}
\end{figure*}

\begin{figure*}
\includegraphics{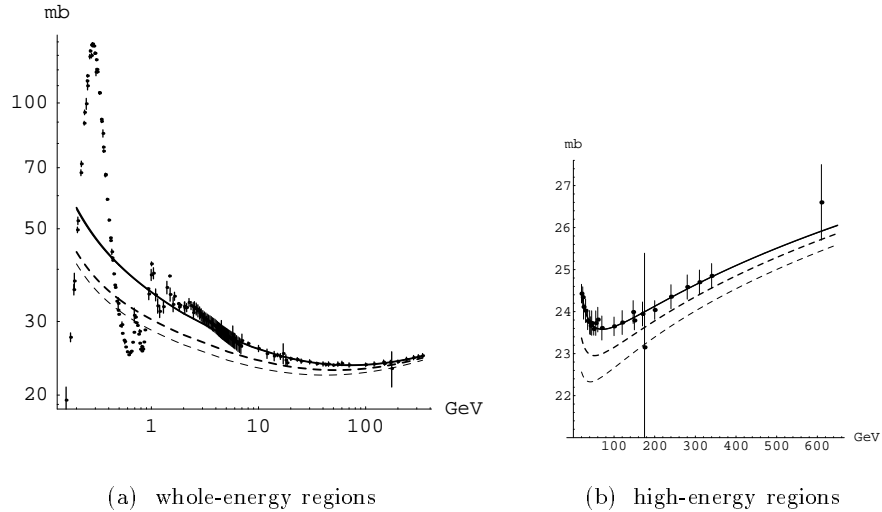}
\caption{\label{fig:wide}Fit to the $\sigma_{\rm tot}^{(+)}$ data 
above 70GeV by the log$^2$ $\nu$ model.
The result in the case of $\alpha_{P^\prime}=0.5$ is shown 
by thick solid line, which overlaps in all energy region 
with the result of $\alpha_{P^\prime}=0.586$ 
shown by thin solid line, and both results cannot be distinguished from each other.
The contribution from Im $R(\nu )$ with $c_2>0$
is shown by thick(thin) dashed line.
A datum at 610 GeV obtained by SELEX collaboration is included in (b). 
Our log$^2\ \nu$ model predicts 25.9mb for 
$\sigma_{\rm tot}^{(+)}$ at 610GeV
which is consistent with their value on $\pi^- N$, 
$(26.6\pm 0.9)$mb .
}
\end{figure*}

\hspace*{-0.4cm}{\bf (2) log$^2$ $\nu$ model}:\ \  
This model has four parameters $c_0$, $c_1$, $c_2$ and $\beta_{P^\prime}$ 
with two constraints FESR(1),(2). 
(So the number of independent parameters is two.)
We again set $N=10$GeV and required both FESR(1) and (2) as constraints.
Then $c_0$, $\beta_{P^\prime}$ are expressed as functions of $c_1$ and $c_2$ as
\begin{eqnarray}
c_0(c_1,c_2) &=& 0.0879-4.94c_1-21.50c_2,\nonumber\\ 
\beta_{P^\prime}(c_1,c_2) &=& 0.1290+7.06c_1+41.46c_2 .\ \ \ 
\label{eq13}
\end{eqnarray}  
We then searched for the fit to 12 data points of $\sigma_{\rm tot}^{(+)}(k)$
above 70GeV. 
The result is shown by thick solid line in Fig.2 .
The best fit in terms of two parameters $c_1$ and $c_2$ led us 
to greatly improved value of ``reduced $\chi^2$," 
$\chi^2/(N_{\rm data}-N_{\rm param})=0.746/(12-2)\simeq 0.075$ for
$c_1=-0.0215<0$ and $c_2=0.00182>0$ which give 
$c_0=0.155$ and $\beta_{P^\prime}=0.0524$. 
This is an excellent fit to the data (see, thick solid line in FIG. 2 (b)).\\

\hspace*{-0.4cm}(\underline{\it $\alpha_{P^\prime}$ dependence})\ \  So far,
we have assumed the intercept of the $P^\prime$ trajectory $\alpha_{P^\prime}(0)$ to be
0.5. The value  $\alpha_{P^\prime}(0)$ is estimated to be 0.586 according to the 
Chew-Frautschi plot, using the universal slope $\alpha^\prime =1/1.15$GeV$^{-2}$
and the mass of $f_2$ to be 1275MeV. Let us check if the results change for this value of 
$\alpha_{P^\prime}(0)$.

Suppose we take $\alpha_{P^\prime}(0)=0.586$ and discuss the two cases, log $\nu$
and log$^2\nu$.\\
{\bf (1) log $\nu$ model}:\ \ We again set $N=10$GeV and expressed both
$c_0,\ \beta_{P^\prime}$ as functions of $c_1$ using FESR (1),(2). We then obtained 
\begin{eqnarray}
c_0(c_1) &=& 0.0817-5.28c_1,\nonumber\\ 
\beta_{P^\prime}(c_1) &=& 0.1238+6.77c_1 .\ \ \ 
\label{eq18}
\end{eqnarray}  
We then searched for the fit to 12 data points of $\sigma_{\rm tot}^{(+)}(k)$ 
above 70 GeV. 
The result is shown by thin solid line in FIG. 1.
The best fit we obtained is $c_1=0.00353$ which gives
$c_0=0.0630$ and $\beta_{P^\prime}=0.148$ with ``reduced $\chi^2$"
$\chi^2/(N_{\rm data}-N_{\rm param})=22.30/(12-1)=2.03$.
So, this model has difficulties again to reproduce the experimental increase
of $\pi p$ total cross sections above 70 GeV 
(see, thin solid line in Fig. 1 (b) ).\\
{\bf (2) log$^2\nu$ model}:\ \ We also set $N=10$GeV and required both 
FESR (1) and (2) as constraints. Then we obtained 
\begin{eqnarray}
c_0(c_1,c_2) &=& 0.0817-5.28c_1-23.50c_2,\nonumber\\ 
\beta_{P^\prime}(c_1,c_2) &=& 0.1238+6.77c_1+39.80c_2 .\ \ \ 
\label{eq19}
\end{eqnarray}  
We again searched for the fit to 12 points 
of $\sigma_{\rm tot}^{(+)}(k)$ above 70 GeV.
The best fit in terms of two parameters $c_1$ and $c_2$ again 
led us to greatly improved value of ``reduced $\chi^2$",  
$\chi^2/(N_{\rm data}-N_{\rm param})=0.750/(12-2)=0.075$
for $c_1=-0.0197<0$ and $c_2=0.00173>0$ which give $c_0=0.145$ and 
$\beta_{P^\prime}=0.0593$. This is again an excellent fit to the data
(see, the caption of Fig. 2).

We have also searched for $\alpha_{P^\prime}(0)=0.543$ (average of 0.5 and 0.586)
and for $\alpha_{P^\prime}(0)=0.642$ due to the Particle Data Group\cite{rf1}. 
We found that the results do not change so much.  

It is remarkable to notice that the wide range of data ($k\geq 5$GeV) 
have been reproduced within the error even in the region where the fit 
has not been made (see Fig. 2 (a) and (b)). 
The results do not change so much for the value of $N$. 
The increase of $\sigma_{\rm tot}^{(+)}$ above 50 GeV
is explained via log$^2$ $\nu/\mu$ $(c_2>0)$ and the decrease between 
$5\sim 50$GeV is explained by log $\nu/\mu$ ($c_1<0$).
It should also be emphasized that the comparison of Fig.~1 (log $\nu$ model)
and Fig.~2 (log$^2\ \nu$ model) clearly indicates the latter model to 
be preferred.
Similar conclusion is also obtained 
by completely different approach\cite{rf11}.


Recently a datum\cite{rf12} for $\pi^- N$ total cross section 
at very high energy ($k$=610GeV)\footnote{
At this energy the difference between $\sigma^{\pi^-p}_{\rm tot}$ and 
$\sigma^{\pi^-n}_{\rm tot}(=\sigma^{\pi^+p}_{\rm tot})$ being estimated 
as $\sim$0.2mb is negligible, and the $\sigma^{\pi^-N}_{\rm tot}$ can be 
regarded as $\sigma^{(+)}_{\rm tot}$). }
was reported by the SELEX collaboration.
Our log$^2\ \nu$ model(log $\nu$ model) with $\alpha_{P^\prime}(0)=0.5$
predicts 25.9mb(24.2mb) for 
$\sigma_{\rm tot}^{(+)}$ at 610GeV
which is consistent(inconsistent) with their value on $\pi^- N$, 
$(26.6\pm 0.9)$mb .
This fact also suggests the validity of the log$^2\nu$ model.

Therefore, we can conclude that our analysis 
in terms of high-energy parameters constrained by the FESR (1),(2) 
prefers the log$^2$ $\nu/\mu$ behaviours satisfying the 
Froissart unitarity bound.
Finally we should add a note that the origin of the log$^2\ \nu$ behaviour 
of the amplitude at high energy is argued to be explained from the effect 
of gluon saturation\cite{rf13}.

\begin{acknowledgments}
One of us (K. I. ) wishes to thank Prof. H. Miyazawa for useful 
conversations. The authors would like to express gratitude to Prof. J. Engelfried
of SELEX collaboration for informing us of their $\pi^-N$ datum at 610GeV.
We also thank Prof. B. Nicolescu for inforimg us of their related works.
   
\end{acknowledgments}

\bibliography{apssamp}

\begin{thebibliography}{99}
\bibitem{rf1} Particle Data Group, D. E. Groom et al., 
Eur. Phys. J. {\bf  C15}, 235 (2000).\\ 
A.~S.~Carroll et al., Phys.~Lett.~{\bf 80B}, 423 (1979); {\bf 61B}, 303 (1976); 
Phys.~Rev.~Lett.~{\bf 33},932 (1974).\\
A.~E.~Brenner et al., Phys.~Rev.~{\bf D26}, 1497 (1982).\\ 
Fong et al., Nucl.~Phys.~{\bf B102}, 386 (1976).\\
Brick et al., Phys.~Rev.~{\bf D25}, 2794 (1982).\\
A.~Firestone et al., Phys.~Rev.~{\bf D14}, 2902 (1976).\\
D.~Ljung et al., Phys.~Rev.~{\bf D15}, 3163 (1977).
\bibitem{rf2} M. Froissart, Phys. Rev. {\bf 123}, 1053 (1961).\\
A.~Martin, Nuovo Cimento {\bf 42}, 930 (1966).
\bibitem{Cudell} J.~R.~Cudell et al., Phys.~Rev.~{\bf D61}, 034019 (2000). 
\bibitem{rf3} K.~Igi, Phys. Rev. Lett. {\bf 9}, 76 (1962).
\bibitem{rf4} G.~F.~Chew, M.~L.~Goldberger, F.~E.~Low and Y.~Nambu,
Phys.~Rev.~{\bf 106}, 1337 (1957).
\bibitem{rf5} K.~Igi and S.~Matsuda, Phys. Rev. Lett. {\bf 18}, 625 (1967).
\bibitem{rf6} A.~A.~Logunov, L.~D.~Soloviev and A.~N.~Tavkhelidze, 
Phys.~Lett.~{\bf 24B}, 181 (1967).
\bibitem{rf7} R.~Dolen, D.~Horn and C.~Schmid, 
Phys. Rev. Lett. {\bf 19}, 402 (1967); Phys.~Rev.~{\bf 166}, 1768 (1968).
\bibitem{rf8} R.~K.~Bhaduri,``Models of the Nucleon," 
Addison Wesley pub., p.134 (1988).
\bibitem{KN75} K.~Kang and B.~Nicolescu, Phys.~Rev.~{\bf D11}, 2461 (1975).
\bibitem{rf11} J.~R.~Cudell et al. (COMPETE collaboration), hep-ph/0107219.
\bibitem{rf12} U.~Dersch et al. (SELEX collaboration),
Nucl.~Phys.~{\bf B579}, 277 (2000). 
\bibitem{rf13} A.~I.~Shoshi, F.~D.~Steffen and H.~J.~Pirner, hep-ph/0202012. 
\end{thebibliography}

\end{document}